\documentclass[12pt]{article}
\usepackage{graphicx}
\begin{document}
\centerline{\bf Simulation of Demographic Change in Palestinian Territories}
\bigskip
\centerline{M. A. Sumour$^1$, A. H. El-Astal$^1$, M. M. Shabat$^2$, and M. A. Radwan$^1$} 
\bigskip
\noindent
$^1$ Physics Department, Al-Aqsa University, P.O.4051, Gaza, Gaza 
Strip,
Palestinian Authority.\\
$^2$ Max Planck Institute for the Physics of Complex Systems,
N\"othnitzer Strasse 38, 01187 Dresden, Germany. \\ 
\bigskip
\{msumoor, a\_elastal, ma.radwan\}@alaqsa.edu.ps, shabat@mail.iugaza.edu 
\bigskip

 
\bigskip

{\bf Abstract}

Mortality, birth rates and retirement play a major role in demographic changes.
In most cases, mortality rates decreased in the past century without noticeable
decrease in fertility rates, this leads to a significant increase in population
growth. In many poor countries like Palestinian territories the number of births
has fallen and the life expectancy increased.

In this article we concentrate on measuring, analyzing and extrapolating the age
structure in Palestine a few decades ago into future.   A Fortran program has
been designed and used for the simulation and analysis of our statistical data. This study of demographic change in Palestine has shown that Palestinians will
have in future problems as the strongest age cohorts are the above-60-year olds.
We therefore recommend the increase of both the retirement age and women
employment.

\bigskip
{\bf Introduction}

During the last two centuries in peaceful rich countries, people lived on average longer and longer, while during the last few decades the number of children born per women during her lifetime has sunken below the replacement rate of slightly above two.  Also in many poor countries the birth rate has decreased and the life expectancy increased \cite{Zekri}.

Mortality, birth rates and retirement have an important effect in demographic changes\cite{Fertility}.  Often, mortality rates decreased in the past century and there is no remarkable decrease in fertility rates, which leads to a significant increase in population growth.  A decrease in the life expectancy is rather rare such as Russian males and some 
African countries.

This study concentrates on measuring, analyzing and studying mortality, birth rates and retirement in Palestine.  Researchers in the field of sociophysics tried to shed lights on the labor market from decreasing fertility rates to find out the major factors that could decrease fertility rates, to find out the major factors that could decrease fertility rates, keep in mind that these factors might differ from a country to another\cite{Zekri,Fertility,Bonkowska,Stauffer1}  

\begin{figure}[hbt]
\begin{center}
\includegraphics [angle=-90,scale=0.5]{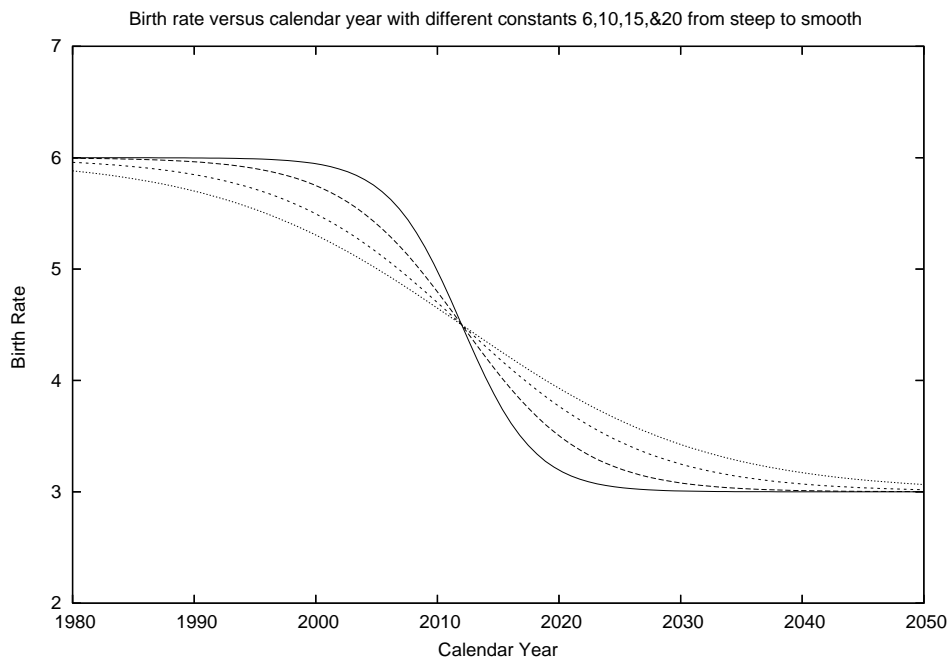}
\end{center}
\caption{ Different birth rate versus calendar year for different constants $6$, $10$, $15$, $20$, which shows the birth rate will be about $3$ after year $2020$}
\end{figure}

\begin{figure}[hbt]
\begin{center}
\includegraphics [angle=-90,scale=0.5]{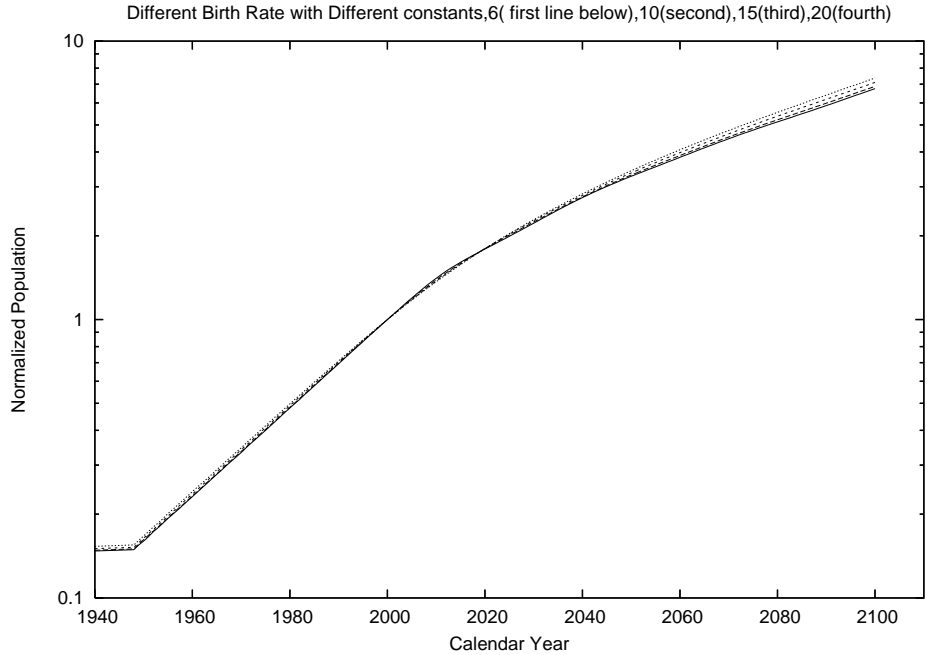}
\end{center}
\caption{Population versus calendar year for different constants 6, 10, 15, \& 20 in the birth rates.}
\end{figure}

\begin{figure}[hbt]
\begin{center}
\includegraphics [angle=-90,scale=0.5]{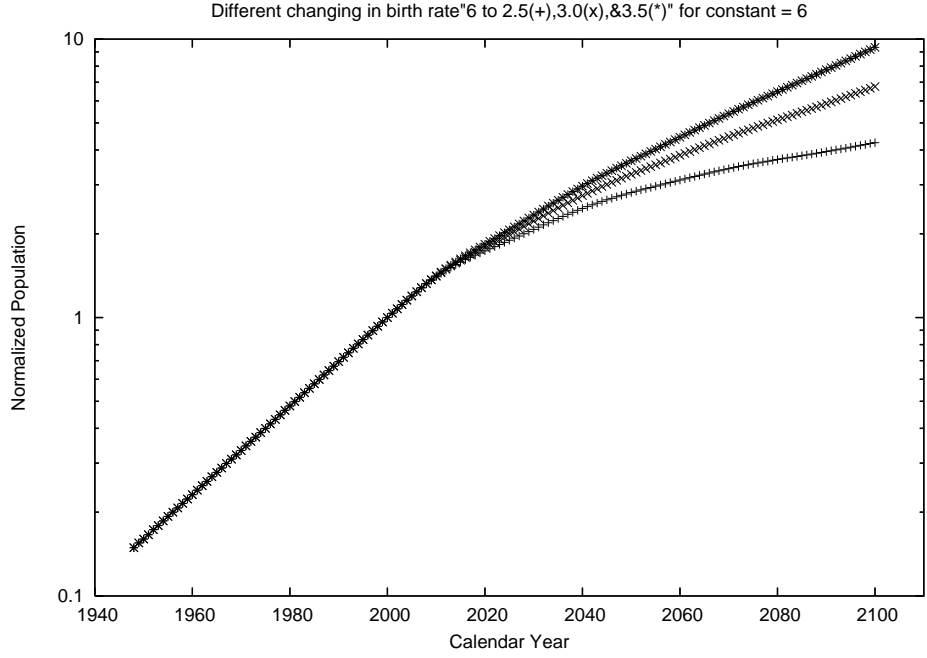}
\end{center}
\caption{Population for different birth rate versus calendar year for  constant  6.}
\end{figure}

\begin{figure}[hbt]
\begin{center}
\includegraphics [angle=-90,scale=0.5]{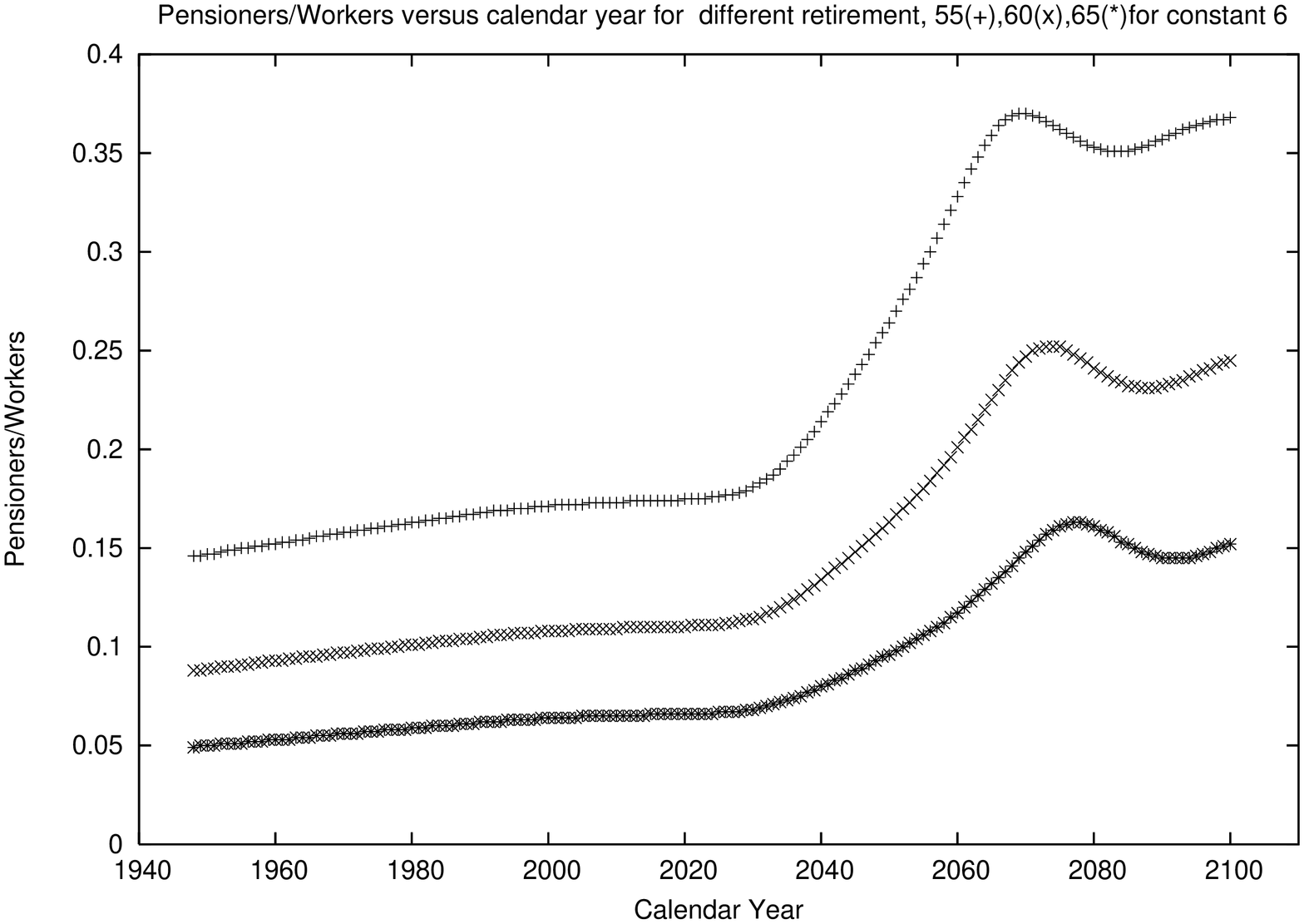}
\end{center}
\caption{Ratio of number of pensioners to number of working age people with different birth rate versus calendar year for different retirement age.}
\end{figure}

\begin{figure}[hbt]
\begin{center}
\includegraphics [angle=-90,scale=0.5]{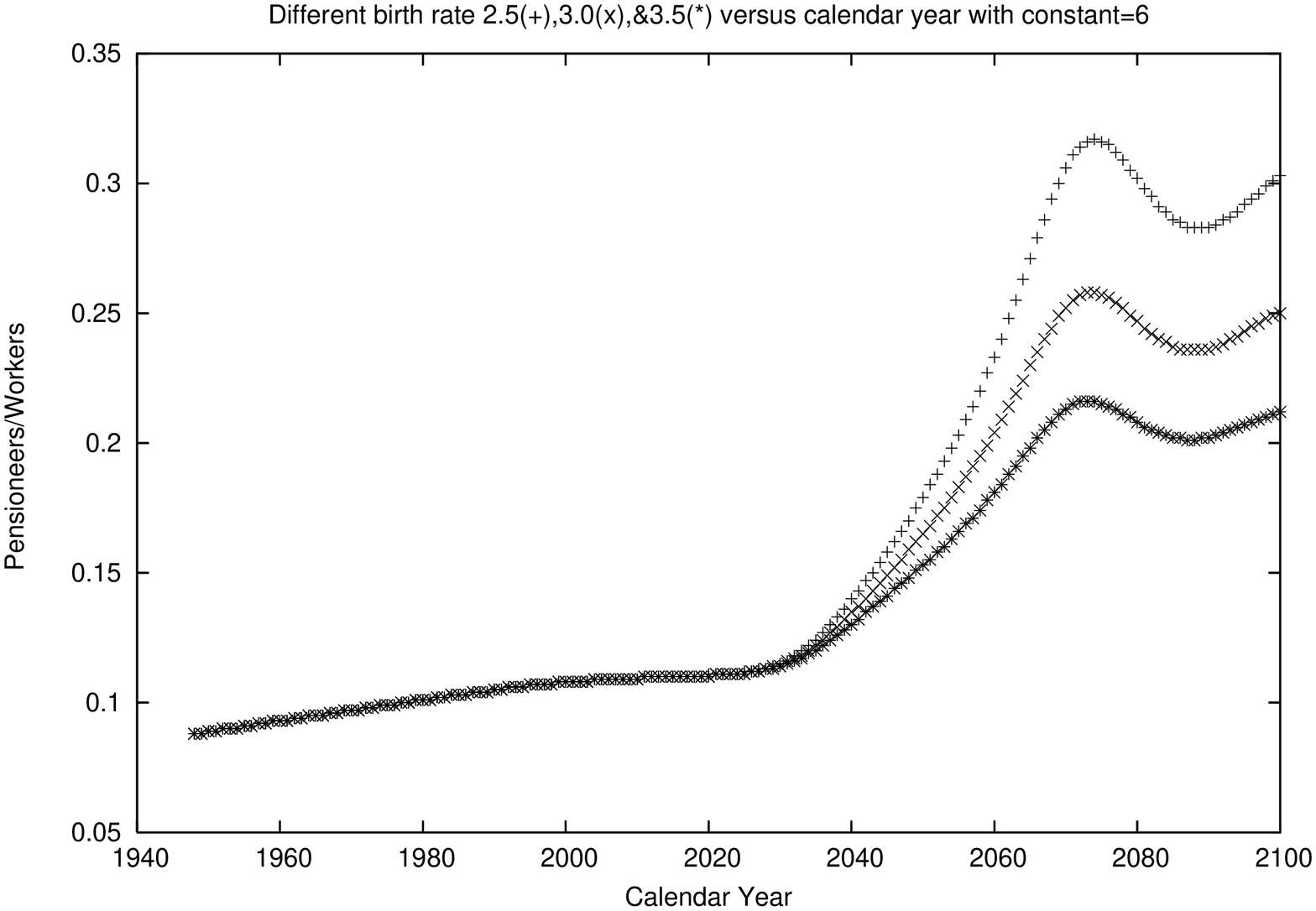}
\end{center}
\caption{ Ratio of number of pensioners to number of working age people with different birth rate 2.5(+), 3.0(x), \& 3.5(*) versus calendar year for  retirement age = 60 years.}
\end{figure}

	Palestinian women now give birth on average to six children in their lives\cite{Fertility}, in Israel less than three, in Algeria more than two, in Spain, Italy, and Germany 1.3 to 1.4, and in Poland and the Czech Republic about 1.2 \cite{Zekri}.
Although fertility rate in the Palestinian Territories is decreasing, it is still high, particularly when it is compared with the worldwide fertility rates. Total fertility rate decreased in Palestinian Territories from 6.77 in 1992 to 6.06 births in 1994 to 5.93 births in 1999.  The fertility rate in West Bank is lower than in Gaza Strip.  The Palestinian Statistics expected the birth rate in 2025 to be about 3.06 \cite{Fertility}.

Our current research deals with Gaza Strip and West Bank. East Jerusalem is not included due to the lack of the statistical data.  The date of our simulation and calculation in this article starts from the year of  Nakba 1948.
 
\bigskip

{\bf Mortality and birth rates}

Now we study the population versus the calendar year by taking the different birth rate with average retirement age $60$, and take the year $2000$ as the reference, so we divide the number of total population by that in the year $2000$, and simulate it to get different values for normalized population. Then we plot the normalized population versus calendar year as presented in figure 2, showing the increase in the total population

According to Azbel\cite{Azbel}, in all different countries and centuries, the probability of humans to survive up to a fixed age is a universal function of the life expectancy; we do not have to apply this universality to yeast cells for the purpose of human demography. 
The mortality function $\mu$ at middle age increases exponentially with age\cite{Stauffer1} :
$$                            \mu(a)  \propto \exp(ba)	 \eqno(1)$$
thus:
$$                            \mu = - d[\ln S(a)]/da	   \eqno(2) $$  
where $S_(a)$ is the number of survivors from birth to age a, is assumed to follow a Gompertz law for adults: 
$$                            \mu =  {\rm Const} \; b \exp[(a- X)b]  \eqno(3) $$ 
since the deviations at young age occur at such low mortalities that they are not relevant if we want to be accurate within a few percent.  The deviations at old age are not yet reliably established and may also be negligible as long as the fraction of centenarians among pensioners is very small \cite{Robine}.

       The Gompertz slope $b$ was assumed to increase linearly with time from   0.06 in 1821 to 0.10 in 1990 and to stay constant thereafter, in contrast to Bomsdorf\cite{Bomsdorf} and Azbel\cite{Azbel}, but in agreement with Yashin et al\cite{Yashin}; see also Wilmoth et al\cite{Wilmoth}.  Instead, the characteristic age $X$  was constant at 103 years until 1990 and then increased each year by 0.01 years to give a rising life expectancy.  Also these deviations from universality are not yet established reliably. 
       Babies are born by mothers of age $21$ to $40$ with age-independent probability.  The average number of children born per women over her lifetime and reaching adult age is assumed to be:
$${\rm Number \; of \; births} =6+(3.06-6)*[1+\tanh((t-2012)/{\rm const})]/2 \eqno(4) $$
 where $t$ is the calendar year and called  iyear in the Fortran program. 
In our simulations we take into account neither immigration nor emigration.
 Immigration and emigration will strongly depend on decisions taken according to the political situations. 
We assume the retirement age is between 55 and 65 years, so we take the central retirement as 60 years. The life expectancy at birth was 72 years in 2000.

\bigskip
{\bf Simulations and results}

Using the Fortran program, the constants a and d are estimated depending on the available Palestinian statistical data.  This estimation was carried out by fitting the theoretical model shown in equations 1-4 with the Palestinian statistical data.  The program can facilitate the calculation and studying of many important aspects such as: a) the overall rise of population, b) the ratio of people above retirement age to people in working age, c) the influence on points a and b of the changing birth rate, d) the influence on a change in retirement age on point b.

In Palestinian Territories the worker is defined as being between 20 years and retirement age is between 55-65 years, on average 60 year.  The birth rate for Palestinian women is decreasing for many different reasons, thus we can simulate these changes for different constants (6, 10, 15 \& 20) in the equation 4, and get figure 1.

But figure 2 proves that the value of constant does not matter much.  So we take the sample for constant = 6,  and the birth rate as changing from 6.0 to 2.5, 3.0, and 3.5 for central retirement age of 60 years, and get figure 3.

We take the ratio of pensioners to workers  for different retirement age 55, 60, and 65 with the birth rate changing from 6.0 to 3.0 as shown in figure 4 using the methods described in references\cite{Stauffer1,Stauffer2,Martins}.

We can also take the ratio of pensioners to workers for retirement age = 60 with the different birth rate (2.5, 3.0, \& 3.5), thus we produce figure 5, using the same methods described in \cite{Bonkowska}, showing that in the year 2030 the effect of birth rate does not matter in the working age people, but after this year it is affected and the number of old people to be supported by working age people will increase drastically, while the working-age fraction decreases.

\bigskip

{\bf Conclusion}

To the best knowledge of the authors, this study is the first one concerning the demographic change in Palestinian Territories. We conclude that Palestinians will have problems similar to Germans  who now retire at an age near 60 years and where around 2030 the strongest age cohorts are near 70 years; Palestinian problems lie ahead only four decades later. Increasing the retirement age, and having more women working, may reduce the problems.

\bigskip

{\bf Acknowledgement:}
The authors thank D. Stauffer for many valuable suggestions and fruitful discussions during the development of this work. 

\bigskip
{\bf Appendix}

Fortran program used in our simulations:

{\small
\begin{verbatim}
       REAL*8 S(0:130),POP(0:130),Q(130),A,B,X,X0,BABIES
       DATA X0/103/,D/0.01/,A/12.0/,B0/0.060/,BMAX/0.10/,BIRTH0/3.0/
     1 ,MENOP/40/,C/.0000/,CONST/6.0/
C      YASHIN X=X0+D*T; GAVRILOV-AZBEL-GOMPERTZ Q/B = A*EXP(B(A-X))
       PRINT *, X0, A, MENOP, B0, C, D,CONST
       B=B0
       X=X0
C      CONST=(2.2-BIRTH0)/150.
       DO 3 IYEAR=1321,2100
       BIRTH = 6 + (3.06-6)*(1+TANH((IYEAR-2012)/CONST))/2
       IF(IYEAR.GE.1841.AND.IYEAR.LE.1990)
     1  B=B0 + (IYEAR-1841)*(BMAX-B0)/150
       IF(IYEAR.GT.1990) X=X0+D*(IYEAR-1990)
       DO 1 IAGE=0,130
         S(IAGE)=DEXP(-A*DEXP(-B*X)*(DEXP(B*IAGE)-1.0D0))
C       S = SURVIVAL PROBABILITY FOR GOMPERTZ LAW 
C       POP = ACTUAL SURVIVORS, CAN BE LARGER THAN ONE
C       Q = MORTALITY FUNCTION CALCULATED FROM S
         IF(IYEAR.EQ.1321) POP(IAGE)=S(IAGE)
  1      IF(IAGE.GT.0) Q(IAGE)=DLOG(S(IAGE-1)/S(IAGE))
       BABIES=0.0D0
       DO 4 IAGE=21,MENOP
  4    BABIES=BABIES+POP(IAGE)*(0.5D0/(MENOP-20))*BIRTH
       POP(0)=BABIES
       DO 6 IAGE=130,1,-1
  6      POP(IAGE)=POP(IAGE-1)*(S(IAGE)/S(IAGE-1))
       WORKER=0.0
       PENSIO=0.0
       TOT =0.0
       EXPECT=0.0
C      NUMBERS OF: WORKERS, PENSIONERS, POPULATION, LIFE EXP. AT 65
        DO 5 IAGE=1,130
C         IF(IAGE.GT.65) EXPECT=EXPECT+S(IAGE)/S(65)
         EXPECT=EXPECT+S(IAGE)/S(0)
         SS=POP(IAGE)
         TOT=TOT+SS
         IF(IAGE.GT.20.AND.IAGE.LE.60) WORKER=WORKER+SS
C5      IF(IAGE.GT.60. OR.IAGE.LE.20) PENSIO=PENSIO+SS
  5    IF(IAGE.GT.60) PENSIO=PENSIO+SS
  3    IF(IYEAR.GT.1947)
     1 PRINT 100,IYEAR,TOT,BIRTH,PENSIO/TOT,PENSIO/WORKER,EXPECT,B,X
100    FORMAT(I5,17F17.1)
       STOP
       END
\end{verbatim}
}

\end{document}